\documentclass{PoS}
\usepackage{amsmath}
\def\be{\begin{eqnarray}}
\def\ee{\end{eqnarray}}
\def\bc{\begin{center}}
\def\ec{\end{center}}
\def\om{\omega}
\def\rmd{{\rm d}}
\newcommand{\lsim}{\stackrel{\scriptstyle <}{\phantom{}_{\sim}}}

\title{Multistrangeness in Heavy-Ion Collisions}

\ShortTitle{Multistrangeness in Heavy-Ion Collisions}

\author{\speaker{E.E. Kolomeitsev}
\\
Matej Bel University in Banska Bystrica, Slovakia\\
E-mail: \email{Evgeni.Kolomeitsev@umb.sk}}

\author{I. Melo\\
University of \v{Z}ilina and  Matej Bel University in Banska Bystrica, Slovakia\\
E-mail: \email{melo@fyzika.uniza.sk }}

\author{B. Tom\'a\v{s}ik\\
Matej Bel University in Banska Bystrica, Slovakia and Czech Technical University in Prague, Czech Republic\\
E-mail: \email{Boris.Tomasik@umb.sk }}

\author{D.N.~Voskresensky\\
        National Research Nuclear University ``MEPhI'', Moscow\\
        E-mail: \email{D.Voskresensky@gsi.de}}
\abstract{
We discuss  strangeness production in heavy-ion collisions
in the broad energy range --- from SIS energies through AGS-SPS-RHIC and upto LHC energies.
On several examples we demonstrate how the strange particle production can reveal information about the collision dynamics and about possible modifications of particle properties in medium. In particular the production of hadrons containing two and more strange quarks, like $\Xi$ and $\Omega$ baryons, or a $\phi$ meson is of interest. We conduct our discussion in the framework of the minimal statistical model, in which the total strangeness yield is fixed by the $K^+$ multiplicity. It is emphasized that in collisions with a small number of produced strange particles, the exact strangeness conservation in each collision event must be explicitly preserved.
}

\FullConference{XXII International Baldin Seminar on High Energy Physics Problems,\\
		15-20 September 2014\\
		JINR, Dubna, Russia}

\begin{document}

\section{Strangeness as a messanger}

Strangeness production in heavy-ion collisions (HICs) enjoys a persistent interest. It is explained by several reasons. First, strangeness is a tag on a hadron, saying that this particular hadron was most probably produced in the course of collision and does not stem from the original nuclei.
Second, strange quarks and anti-quarks prefer to reside in different hadronic species.
A strange quark prefers to be in baryons: so there is two kaon $\bar{K}$ states for several baryonic states $\Lambda$, $\Sigma$ ,$\Xi$, $\Omega$. Oppositely, the anti-strange quark  can only be in mesons, like $K$ and its heavier resonances. This difference leads to distinct interactions of strange and anti-strange subsystems with a non-strange baryonic environment. The strange subsystem couples more strongly than the anti-strange one. As argued in Ref.~\cite{Ko83,KVK96,Greiner91} this could leads to strangeness/anti-strangeness separation in baryon-rich matter formed in HIC at energies from SIS to middle RHIC energies.
Third, strangeness is conserved  in strong interactions. Therefore the strange and anti-strange particles are created in elementary processes only pairwise. This renders high thresholds for strange particle production, making yields of strange particles in HIC sensitive to possible in-medium effects.

In spite of the common enthusiasm about strangeness in HIC, such studies are complicated by several factors. First of all, cross sections of reactions with the strangeness production are poorly known, especially close to the reaction threshold. Experimentally accessible are only reactions with the $pp$, $np$ and $\pi p(n)$ entrance channel, as those currently studied, e.g., by HADES collaboration at GSI~\cite{HADES} and by ANKA, HIRES, and TOF collaborations at COSY~\cite{COSY}. Without reliable experimental information about reactions with the $nn$ initial state we have to rely on the isospin symmetry relations, making {\it ad hoc} assumptions about internal mechanisms of the reactions~\cite{Fischer}.  Second, interactions of strange particles with non-strange ones and among themselves cannot be fully constrained. There are limited data in the channels with the total strangeness $S=\pm 1$, like a hyperon-nucleon and anti-kaon--nucleon scattering.
Finally, the properties of strange particles and their interactions in dense and hot nuclear matter are far from being firmly established.

Against all odds mentioned above a satisfactory description of the kaon and $\Lambda$ productions in HICs has been reached in various transport~\cite{Fuchs04,Hartnack12} and statistical~\cite{KVK96,Cleymans} approaches.

There remains however several open issues with the production of particles containing several strange quarks, like $\Xi$ and $\Omega$ baryons and $\phi$ mesons, which we would like to discuss in the present talk.

\section{Strange particles in medium}\label{sec:medium}

Medium effects prove to be important for description of particle production in HIC at SIS energies~\cite{Revs,Voskre-HIC}. Particularly, strange particle yields are strongly influenced by them~\cite{Fuchs04,Hartnack12,KVK96,LLB97,BratCass}. For hyperons and nucleons the in-medium modification of the energy spectrum of particle $a$ is often effectively parameterized in terms of scalar $S_a$ and vector $V_a$ potentials $ E_a(p)=\sqrt{m_a^{*2}+p^2}+V_a$, where the scalar potential enters the spectrum through the effective mass $m_a^* =m_a+S_a $. Description in terms of the $S_a$ and $V_a$ potentials is typical for relativistic mean-field (RMF) models, cf.~\cite{Fuchs04,KV05}.  The nucleon potentials determined in~\cite{KV05} are  $S_N\simeq-190~{\rm MeV}\rho_B/\rho_0$ and $V_N\simeq + 130~{\rm MeV}\rho_B/\rho_0$, here
$\rho_B$ is the baryon density and $\rho_0=0.16/{\rm fm}^3$ is the nuclear saturation density. The same potentials could be also used for $\Delta$: $S_\Delta\simeq S_N$, $V_\Delta\simeq V_N$.

One usually relates the hyperon potentials to the nucleon ones, $V_H=\alpha_H V_N$ and
$S_H=\beta_H S_N$. The parameter $\alpha_H$ one choose according to the number of non-strange quarks in the hyperon, $\alpha_\Lambda=\alpha_\Sigma=2\alpha_\Xi=2/3$.
The parameter $\beta_H$ is, then, chosen such that the optical potential of a hyperon in nuclear medium at saturation $U_H=S_H(\rho_0)-V_H(\rho_0)$ agrees with the empirical information from
the analysis of hypernuclei:  $U_\Lambda=-27$~MeV~\cite{Hashimoto06}, $U_\Sigma=+24$~MeV~\cite{Dabrowski99}, and $U_\Xi=-14$~MeV~\cite{Khaustov00}.

\begin{figure}
\bc
\includegraphics[width=9cm]{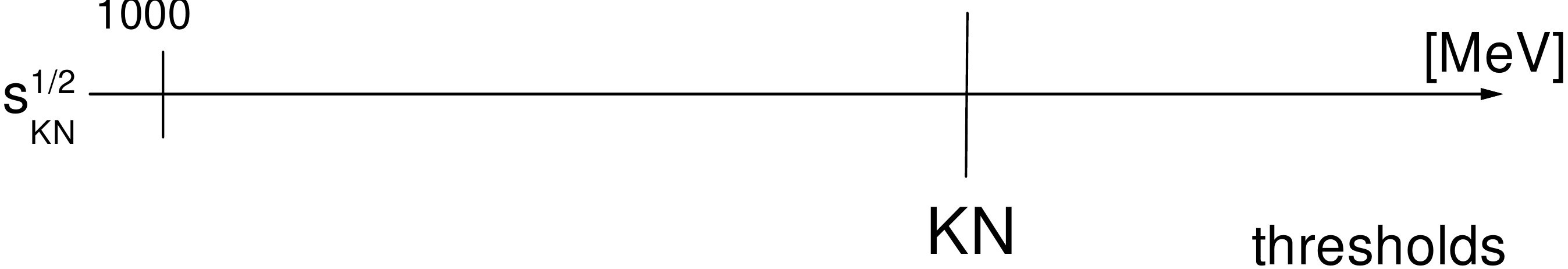} \\[7mm]
\includegraphics[width=9cm]{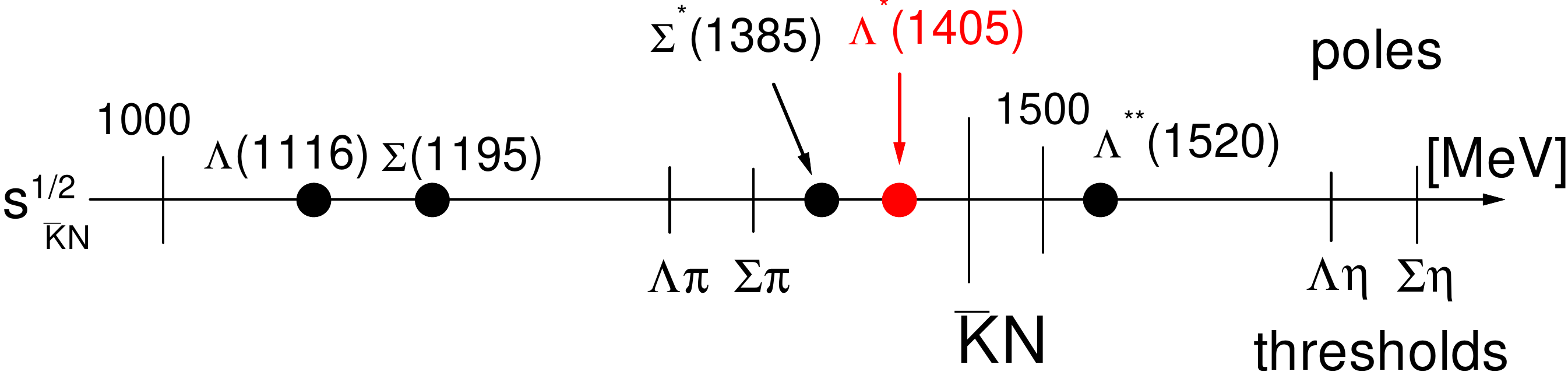}
\ec
\caption{World of kaon-nucleon (upper panel) and anti-kaon--nucleon (lower panel) scattering. Vertical dashes show thresholds and fat dots correspond to poles.}
\label{fig:meson-map}
\end{figure}

In the sector of strange mesons the situation is more controversial.
In Fig.~\ref{fig:meson-map} we depict maps of the worlds of $KN$ and $\bar{K}N$ scattering.
The $KN$ interaction looks structureless without any resonances and elastic up to rather high energies. Mislead by this simplicity, one frequently uses the leading-order chiral Lagrangian to write the kaon self-energy as $\Pi_K(\om,\vec{q}\,)=\Pi_K^{\rm (LO)}(\om)= (3\om/4 +\Sigma_{KN})\rho/f^2$, which determines the kaon spectrum as a solution of the Dyson equation, $\om^2-\vec{q}\,^2-m_K^2- \Pi_K(\om,\vec{q}\,)=0$, with $m_K$ being the kaon mass, $f\simeq 92$~MeV standing for the pion decay constant and the kaon-nucleon $\Sigma$-term $\Sigma_{KN}\simeq 350$~MeV. However, in Ref.~\cite{LKK} it was shown that if the kaon self-energy
is evaluated in terms of the real parts of the s- and p-wave kaon--nucleon scattering amplitudes
obtained from the non-perturbative solution of the Bethe-Salpeter scattering equation~\cite{LK02}
with the kernel given by the chiral perturbation theory, then the kaon self-energy can be parameterized as $\Pi_K^{\rm (NP)}(\om,\vec{q}\,) = \big(1.1\,m_K-\om+0.2\,\vec{q}\,^2/m_K\big) 46.8\,{\rm MeV}  \rho/\rho_0$\,. We see
the scalar and vector terms in $\Pi_K^{\rm (LO)}$ and $\Pi_K^{\rm (NP)}$ have opposite signs. Thus, the chiral Lagrangian does not correctly describe the kaon--nucleon scattering process if evaluated in perturbation theory.

The world of antikaon--nucleon scattering is much richer: it is a coupled-channel system as the $\bar{K}N$ scattering is inelastic already on the threshold due to the coupling to $\pi\Lambda$ and $\pi\Sigma$ channels. There is a bunch of resonances, among which the s-wave $\Lambda(1405)$ resonance dominates the $\bar{K}N$ scattering in the isospin-0 channel close to the threshold. In the p-wave, the resonance $\Sigma(1385)$ determines the isospin-1 scattering amplitude.
The $\bar{K}N$ interaction is strongly attractive right below the threshold this would lead in medium to a shift of the $\bar{K}$ spectral density for $\vec{q}=0$ towards lower energies. There, kaonic modes can couple strongly to the hyperon--nucleon-hole modes ($HN^{-1}$)~\cite{KVK96} built by ground-state hyperons $\Lambda(1116)$ and $\Sigma(1195)$. The complicated  interplay of anti-kaons and hyperon-resonance propagation in medium asks for self-consistent calculations~\cite{LKorp02,Tolos06}. There are attempts to implement the  anti-kaon spectral functions in transport calculations~\cite{Cass-Tolos}.
As argued in Refs.~\cite{Ko83,KVK96} anti-kaons leave the fireball created in a HIC at the last stage of it evolution, when the baryon density is rather low $0.5\rho_0\lsim \rho_B\lsim\rho_0$. Therefore instead of the spectral function one frequently uses an effective scalar potential changing the $\bar{K}$ mass as $m_K^*=m_K+U_{\bar{K}}\rho_B/\rho_0$ with the parameter
$U_{\bar{K}}=-(70\mbox{--}150)$~MeV.

\section{Minimal statistical model for strange particles}

We assume that in HIC a thermalized nuclear system (a fireball) is formed. This hot and dense system participate in a hydrodynamical expansion, which lasts until the moment of freeze-out characterized by the values $\rho_{B,\rm fo}$ and $T_{\rm fo}$. Henceforth,
in-medium particle thermal momentum distributions become distributions of free-streaming particles. To simplify, we assume the fireball to be spatially uniform and characterized by a time-dependent temperature $T(t)$, baryon density $\rho_B (t)$ and volume $V(t)$.
Because of the high production thresholds the strange particles are most efficiently produced at the early hot and dense stage of the fireball evolution. At SIS-SPS energies the fireball is baryon rich, and antistrange particles (kaons, $K^+$ and $K^0$, mesons) have longer mean free paths than strange particles (antikaons and hyperons). Therefore kaons can easily move off the production point, and either leave the fireball immediately or first thermalize via elastic kaon-nucleon scattering and then leave at some intermediate stage. Because the population of strange particles is very small, even if the kaon stays in the fireball for a while, there is little chance that it meets antikaons or hyperon and is absorbed by them. Thus, in the
course of collision the amount of strangeness of the fireball grows. The accumulated strangeness is redistributed among various hadronic species, like $K^-$, $\bar K^0$, $\Lambda$, $\Sigma$, $\Xi$, $\Omega$ and their resonances, which are released at the fireball breakup. (At highest RHIC and LHC energies the strange antibaryons will also contribute.) Thus, the kaon yield measured experimentally or calculated theoretically in some hadro-chemical models can be used to normalize the abundance of strange particles.

We will demonstrate how this approach works in several examples given below.

\subsection{Strangeness at the SPS energy scan. $K^+/\pi^+$ horn}

The excitation function of the ratio $\langle K^+\rangle / \langle \pi^+ \rangle$ which exhibits a sharp peak (a ``horn'') at projectile energies of 30~$A$GeV was intriguing experimental result obtained in the energy scan performed at the CERN SPS~\cite{CERN-data}. Statistical models were  not able to reproduce the observed sharpness of the peak~\cite{Cleymans:2004hj} unless the very broad $\sigma$ meson resonance was included~\cite{APBMJ09}.
The excitation functions of $\langle K^+\rangle / \langle \pi^+ \rangle$, $\langle K^- \rangle/\langle \pi^-\rangle$ and $\langle \Lambda \rangle / \langle \pi \rangle$ are difficult to reproduce in hadronic transport codes either.
In Ref.~\cite{KT05} we calculate relative abundances of strange particles in HIC using a hadronic kinetic model for strangeness production. Being interested only in the ratios of total yields it is enough to study the evolution of the spatially averaged densities of individual species only. The density of kaon is then determined by the equaiton
\begin{equation}
\label{Kprod}
\frac{\rmd \rho_K}{d\tau} = \rho_K \Big( -\frac{1}{V}\, \frac{\rmd V}{\rmd\tau}\Big) + \mathcal{R}_{\rm gain}-
\mathcal{R}_{\rm loss}\,,\quad
\mathcal{R}_{\rm gain}= \sum_{ij}  \frac{\langle v_{ij}\sigma^+_{ij} \rangle }{1+\delta_{ij}} \rho_i\, \rho_j
\, ,
\end{equation}
where $V$ is the volume of the system, $\tau$ is the time in the co-moving frame.
The first term on the right-hand-side  is for the density change due to the fireball expansion. The term $\mathcal{R}_{\rm gain}$ is the kaon production rate with
the summation running over non-strange hadron species with densities $\rho_i$ and the
$\langle v_{ij}\sigma^+_{ij} \rangle$ standing for the reaction cross section averaged with the relative velocity of the colliding particles over the relativistic Boltzmann distributions.
The most important reaction channels with $\pi B$, $BB$ ($B=N,\Delta$) and $MM$ ($M=\pi,\rho$) entrance channels are included.  The last term, $\mathcal{R}_{\rm loss}$, on the right hand side of (\ref{Kprod}) represents the rate of kaon annihilation processes. As demonstrated in~\cite{KT05} for AGS--SPS energies, $\mathcal{R}_{\rm loss}\ll \mathcal{R}_{\rm  gain}$ and can be neglected. Equation (\ref{Kprod}) is integrated starting from some initial  kaon density due to primordial kaon production in collisions of incident nucleons. It was estimated  from a compilation of kaon production data in nucleon--nucleon collisions~\cite{initkaon}.

\begin{figure}
\begin{center}
\includegraphics[width=9cm]{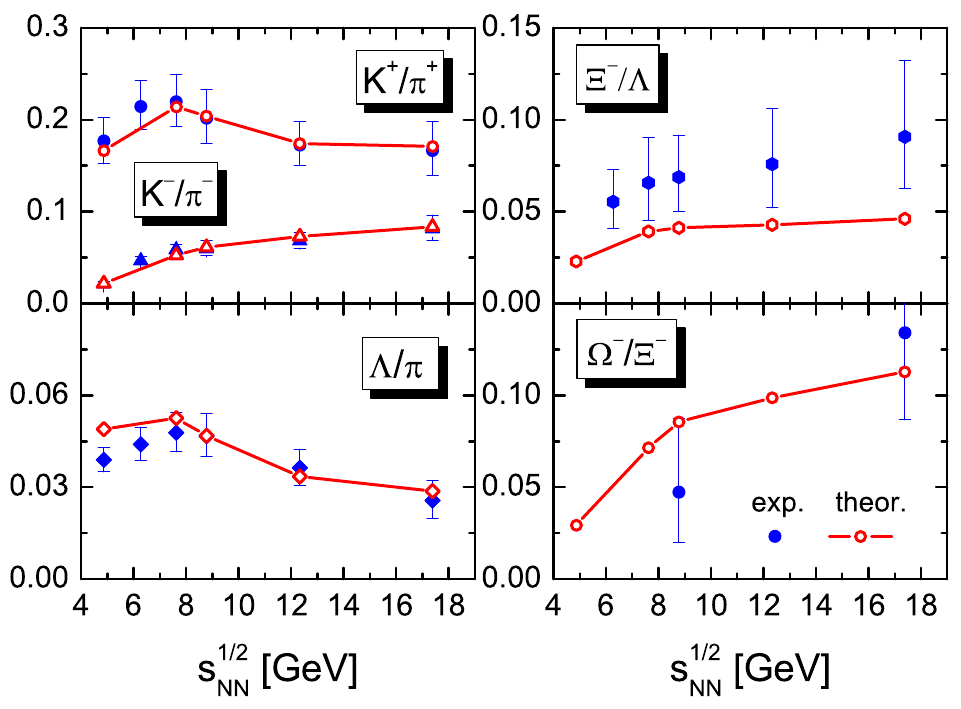}
\end{center}
\caption{Excitation functions of the strange particle ratios calculated in~\cite{KT05} in comparison with the data~\cite{CERN-data,Lung}.}
\label{fig:SPS-ratios}
\end{figure}

For the evolution of the energy density and the baryon density we use parameterizations consisting of an initial accelerating period and a later scaling expansion with power-law dependence of the density on time, whereby the evolution of the baryon density determines actually the expansion of the fireball volume. By exploring a range of parameters we can vary the evolution between the well-known Bjorken and Landau hydrodynamic solutions of the fireball expansion.

The input parameters are the total fireball lifetime and the initial energy density. They, together with the final state densities, determine the whole evolution scenario. The energy density and number densities in the final state are chosen so that they correspond to the values extracted within the statistical model~\cite{Becattini04} from the chemical freeze-out fit of the experimental data.

Solving Eq.~(\ref{Kprod}) we obtain the total amount of anti-strangeness (S=+1) produced in the fireball during its expansion. Species with negative strangeness must balance the total strangeness of the system to zero. Reactions which just swap the strange quark between them are quick. Therefore, we assume that these species are in chemical equilibrium with respect to each other, while the strangeness-weighted sum of their densities is given by the requirement of
strangeness neutrality.

The resulting particle ratios are depicted in Fig.~\ref{fig:SPS-ratios}.
Our hadronic kinetic model~\cite{KT05} is able to describe satisfactorily the excitation function of the ratios $K^+/\pi^+$, $K^-/\pi^-$, and $\Lambda/\pi$. However, the ratio $\Xi^-/\Lambda$ remain underestimated in the entire range of collision energies. This can indicate, perhaps, that the $\Xi$ baryons decouple from the fireball at an earlier stage of the fireball evolution at an higher temperature. In Ref.~\cite{KT05} we have not included in-medium effects neither for hyperons nor for mesons, which are not important at the fireball temperature characteristic for this energy range.

\subsection{$\phi$ production}

The study of $\phi$ meson production in HICs provides complementary information on collision dynamics and in particular on the evolution of the strange subsystem.  Interactions of $\phi$ mesons, consisting mainly of $s$ and $\bar{s}$ quarks, with non-strange hadrons are suppressed due to the Okubo-Zweig-Iizuka rule. Hence, amplitudes of OZI-forbidden reactions are typically orders of magnitude smaller than those of OZI-allowed ones.
Since the OZI suppression weakens the $\phi$ production only in the ordinary hadronic matter and
would be lifted in the quark-gluon medium, the strong, order of magnitude, enhancement of the $\phi$ yield was proposed in~\cite{Shor85} as a signal of the quark-gluon plasma formation.
Experiments at AGS and SPS energies had indeed indicated an enhanced $\phi$ yield albeit to a lesser degree~\cite{phi-AGS}. In~\cite{KoSa} the enhancement was explained by contributions from the OZI-allowed process with strangeness coalescence $K\bar{K}\to \phi\rho$ and $K\Lambda\to\phi N$, provided the $\phi$ meson mass is strongly reduced in medium. Surprisingly strong enhancement of the $\phi$ yield was observed at the beam energies about 2~GeV per nucleon~\cite{phi-FOPI}. Such a large $\phi$ abundance cannot be described by the transport model~\cite{Kampfer02} where $\phi$s are produced in OZI-forbidden reactions $BB\to BB\phi$ and $\pi B\to \phi B$ ($B= N,\,\Delta$) with the dominant contribution from pion-nucleon reactions.
Note that the strangeness coalescence process could not contribute much at these energies, since kaons have a long mean free path and most likely leave the fireball right after they are created.

In~\cite{KB09} we proposed a new mechanism of the $\phi$ production on strange particles,
$\pi Y\to \phi Y$\,, $\bar{K} N\to \phi Y$ with $Y=\Lambda\,,\, \Sigma$\,. In contrast to the coalescence reaction, here the strangeness does not disappear from the system and plays the role of a catalyser. The presence of $K$ mesons is unnecessary.  It was found~\cite{KB09} that the catalytic reactions can be operative if the maximal temperature in nucleus-nucleus collisions is larger than 130~MeV and the collision time is larger than 10~fm.

\begin{figure}
\includegraphics[width=4.7cm]{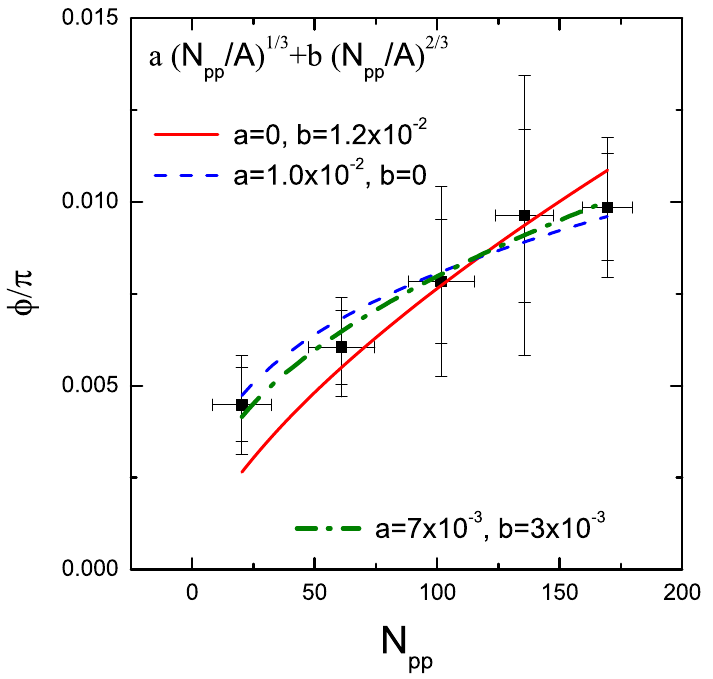}
\includegraphics[width=4.7cm]{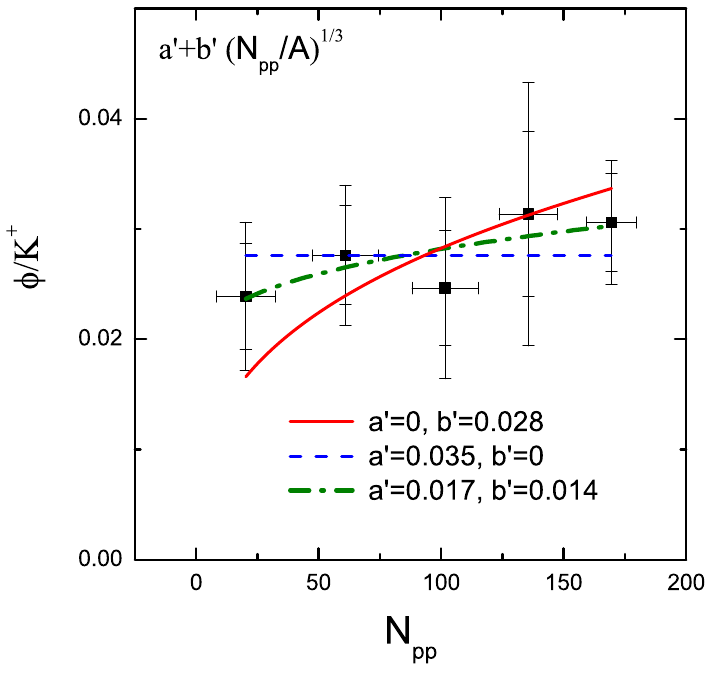}
\includegraphics[width=4.7cm]{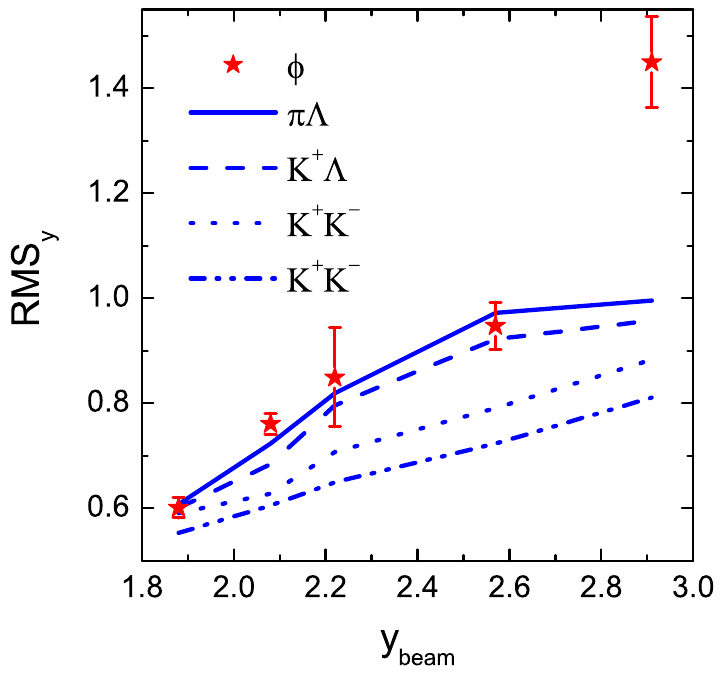}
\caption{
Observables in $\phi$ production indicating the link among $\phi$ mesons and hyperons.
Left and middle panels: centrality dependence of ratios $\phi/\pi$ and $\phi/K^+$.
Data points are from~\cite{AGS-Back04} correspond to Au+Au collisions at 11.7\,$A$GeV. Right panel: width of the $\phi$ rapidity distributions
in PbPb collisions versus the beam rapidity~\cite{NA49-Alt08}. Curves show the rapidity widths
calculated from  rapidity widths of reactant species.
}
\label{fig:phi}
\end{figure}

If the $\phi$ production in HIC is indeed promoted by other strange particles one could expect
correlations among them to be seen in observables.  The centrality dependence of the $\phi$
production was argued in~\cite{KB09} could be such a one. The number of produced $\phi$ mesons
can be estimated as $N_\phi\sim a_{\rm conv} N_{pp}^{4/3}+a_{\rm cat}\, N_{pp}^{5/3}$, where
$N_{pp}$ is the mean number of projectile participants and the first term corresponds to conventional processes like $\pi N\to N \phi$, and the second term is due to the catalytic reactions. The extra power $N_{pp}^{1/3}$ appears in this term because  the number of strange particles in the fireball scales with the fireball expansion time $t_0$ which in turn is proportional to the system size $l\propto V^{1/3} \propto N_{pp}^{1/3}$ for the case of ideal hydrodynamics. For the experimental ratios measured in~\cite{AGS-Back04} for Au+Au collisions at 11.7\,$A$GeV we find
$
{N_\phi}/{N_\pi}\sim a\,n_{\rm pp}^{1/3}+b\, n_{\rm pp}^{2/3}$\,,
and
${N_\phi}/{N_{K^+}}\sim a' + b'\, n_{\rm pp}^{1/3}\,,
$
where $n_{\rm pp}=N_{\rm pp}/A$, $a\,,a'$ and $b\,,b'$ parameterize the relative strength of conventional and catalytic processes, and $A$ is the number of nucleons in the colliding nuclei.
Fits of the experimental data by these two expressions are presented in Fig.~\ref{fig:phi} (left and middle panels). We see that the best fits are reached when both parameters are activated (dash-dotted curves).

A link between $\phi$ meson and hyperons could be also seen in particle rapidity distributions.
If, following Ref.~\cite{NA49-Alt08}, we assume that the rapidity distributions of particles stay unchanged after some initial stage when nuclei are passing through each other, then the rapidity distribution of $\phi$ mesons produced in reaction $1+2 \to \phi + X$ can be approximated as the product of rapidity distributions of colliding particle species 1 and 2.
In Fig.~\ref{fig:phi} (right panel) we show the experimental
root-mean-square of the rapidity distributions of $\phi$s produced in PbPb collisions~\cite{NA49-Alt08} in comparison with the  rapidity distributions width for $\phi$s stemming from $K^+K^-\to \phi$, $\pi\Lambda\to \phi\Lambda$ and $K^+\Lambda \to N\phi$ reactions.
We observe that reactions involving hyperons produce the distributions width in better agreement with the experimental results.

\section{Rare strangeness: $\Xi$ puzzle at GSI energies}

We turn now to HIC at energies close to the strangeness production threshold, like those studied by FOPI and HADES collaborations at GSI. Now the strangeness production is really a very rare event and we have to respect the strangeness conservation on event by event basis.
Depending on the number of $s$ quarks remaining in the fireball (recall, $\bar{s}$ quarks leave the fireball in form of $K$ mesons at early stage of the collision), various strange hadrons and their combinations can be observed in the final state. Hence the chemical equilibrium conditions are different for the events with different numbers of produced kaons.
To handle this complication we divide the totality of events with strangeness production into
classes of events with one, two, three and so on, $s\bar{s}$ quark pairs created.
We call them $n$-kaon events and denote the probability of creation of exactly $n$ strange quark pairs as $P_{s\bar s}^{(n)}$. The probability is determined by the Poisson distribution
$P_{s\bar s}^{(n)} = W^n \,e^{-W} /n!$\,, with the integral probability of the creation of one $s\bar{s}$ pair given by $W = \lambda\, V_{\rm fo}^{4/3}$
where $V_{\rm fo}$ is the fireball freeze-out volume determined by the freeze-out baryon density  $\rho_{B,{\rm fo}}$ and the overlap volume of the initial colliding nuclei. So, for a symmetrical collision of two nuclei with the atomic number $A$ we can write
$V_{\rm fo}(b) = 2\,A\,F(b/b_{\rm max})/\rho_{B,{\rm fo}}$\,, where $F(x)$ is the overlap function parameterized, e.g., in Ref.~\cite{Gosset77}. The parameter $\lambda$ can be fixed by the total multiplicity of $K^+$ mesons observed in an inclusive collision, $\mathcal{M}_{K^+}=\sum_{n} \langle n P_{s\bar{s}}^{(n)}\rangle/(1+\eta)$. Here the parameter  $\eta=(A-Z)/Z$ quantifies the braking of the isospin symmetry in HIC. (For example in collisions Ar+KCl studied by HADES $\eta\simeq 1.14$.)
The anglular brackets stand for the averaging over collision impact parameter
$ \langle \dots\rangle=2 \int_0^{b_{\rm max}} \rmd b\, b\, (\dots)/b^2_{\rm max}$\,,
where the integration over the impact parameter runs up to the maximal possible value $b_{\rm max}$. As a result for the parameter $\lambda$ we find $\lambda=(1+\eta)\,\mathcal{M}_{K^+}/\langle V_{\rm fo}^{4/3}\rangle$.

The statistical probability that strangeness will be released at freeze-out in a hadron of type $a$ with the mass $m_a$ is given by the standard Gibbs' formula
\begin{align}
P_a = z_S^{s_a}\,V_{\rm fo}\, p_a= z_S^{s_a}\,V_{\rm fo}\, \nu_a\, e^{B_a\frac{\mu_{B,{\rm fo}}}{T_{\rm fo}}} f(m_a,T_{\rm fo}),
\quad
f(m,T) = \frac{m^2\, T}{2\pi^2} K_2\left(\frac{m}{T}\right ),
\label{h-density}
\end{align}
where $B_a$ is the baryon number of the hadron, the degeneracy factor $\nu_a$ is determined by
the hadron's spin $I_a$ and isospin $G_a$ as $\nu_a=(2\, I_a+1)\,(2\, G_a+1)$\,, and $K_2$ is the MacDonald function. The baryon chemical potential at freeze-out is determined by
$
 \mu_{B,{\rm fo}}\!\simeq\! -T_{\rm fo}
\ln\big(4\big[f(m_N,T_{\rm fo})+4f(m_\Delta,T_{\rm fo})\big]/{\rho_{B,{\rm fo}}}\big),
$
where $m_N$ is the nucleon mass. $\Delta$ isobars are also treated as stable particles with
the mass $m_\Delta=1232$~MeV, and the small contributions of heavier resonances, hyperons, and anti-particles are neglected.

The quantity $z_S$ in Eq.~(\ref{h-density}) is a normalization factor. It is related to the probability of finding one $s$ quark in the hadron $a$. The factor $z_S^{s_a}$ follows from the requirement that the sum of probabilities for production of different strange species and their combinations, which are allowed in the finale state, is equal to one. The factor $z_S^{s_a}$ depends on how many strange quarks are produced. Hence, it is different in single-, double- and triple-kaon events. Therefore, we introduce the notation $ P_a^{(n)} = (z_S^{(n)})^{ s_a}\,V_{\rm fo}\, p_a$\,, where the superscript $n$ indicates to which class of events this probability and the $z_S$ factor belong:\\
(i)~In \emph{a single-kaon event} one $s$-quark can be released  as $\bar{K}$, $\Lambda$ or $\Sigma$. Hence, the normalization condition for the probabilities (\ref{h-density}) reads
$ z_S^{(1)}\,V_{\rm fo}\, (p_{\bar{K}}+p_{\Lambda}+p_{\Sigma})=1. $
The multiplicity $M_a^{(1)}$ of strange hadrons of type $a=\{\bar{K}, \Lambda, \Sigma\}$ produced in such single-kaon events is given then by
$
M_{a}^{(1)}= g_a \, P_{s\bar{s}}^{(1)}\,z_S^{(1)}\, V_{\rm fo}\, p_a
$\,.
The isospin factor $g_a$ takes into account the asymmetry in the yields of particles with various isospin projections induced by the global isospin asymmetry of the collision $\eta$.  For $\eta=1$ this factor reduces to the standard $1/(2\, G_a+1)$\,.\\
(ii)~In \emph{a double-kaon event} there can be one $\Xi$ baryon and all possible combinations of kaon and hyperon pairs. For double-kaon events the normalization condition is determined by equation
$z_S^{(2)2}\,V^2_{\rm fo}\, (p_{\bar{K}}+p_{\Lambda}+p_{\Sigma})^2 + z_S^{(2)2}\,V_{\rm fo}\, p_{\Xi}=1.
$
The factors 2, which would appear here at the cross
terms after opening the brackets, reflect the number of combinations with which two $s$ quarks can be released as a given combination of hadrons, for example, $\bar{K}\Lambda$, $\bar{K}\Sigma$ and $\Sigma\Lambda$.
Now the multiplicity of the particle $a$ with one $s$ quark produced given by
$
M_a^{(2)} =g_a\, 2\,P_{s\bar{s}}^{(2)}\,P^{(2)}_{a}\sum_{b} P^{(2)}_{b}
$
where $a,b=\{\bar{K}, \Lambda, \Sigma\}$. We take here into account that the hadron $a$ can be produced in pair or in various combinations with other strange hadrons. In both cases the quark combinatoric factor 2 has to be included. The multiplicity of produced $\Xi$ baryons is $ M_\Xi^{(2)} = g_\Xi\,P_{s\bar{s}}^{(2)}\,\,P^{(2)}_{\Xi}$ \,.\\
(iii)~In the very rare \emph{triple-kaon event}, when three $K^+$ mesons are produced, the normalization condition is
$ z_S^{(3)3}\,V^3_{\rm
fo}\,(p_{\bar{K}}+p_{\Lambda}+p_{\Sigma})^3
+ 3\,z_S^{(3)3} V^2_{\rm fo}(p_{\bar{K}}+p_{\Lambda}+p_{\Sigma})
p_\Xi +z_S^{(3)3}V_{\rm fo} p_{\Omega}=1$.
Factors 3 hidden in this relation show in how many different ways three $s$ quarks can be distributed between three hadrons (e.g., $\bar{K}\Lambda\Lambda$,  $\bar{K}\bar{K}\Lambda$) or two hadrons (e.g., $\Xi\bar{K}$, $\Xi\Lambda$). The multiplicity of hadrons with one $s$ quark is
$M_a^{(3)}=g_a\,3 P_{s\bar{s}}^{(3)} P^{(3)}_a\,
[\big(P^{(3)}_{\bar{K}}+P^{(3)}_{\Lambda}+P^{(3)}_{\Sigma}\big)^2 +  P^{(3)}_\Xi]
$.
For hadrons with two $s$ quarks we find
$M_\Xi^{(3)}=g_\Xi\,3P_{s\bar{s}}^{(3)} \, P^{(3)}_\Xi\,(P^{(3)}_{\bar{K}}+P^{(3)}_{\Lambda}+P^{(3)}_{\Sigma})
$
and for the $\Omega$ baryon
$
M_\Omega^{(3)}=P_{s\bar{s}}^{(3)}\, P^{(3)}_\Omega
$.

\begin{figure}
\centering
\includegraphics[width=7cm]{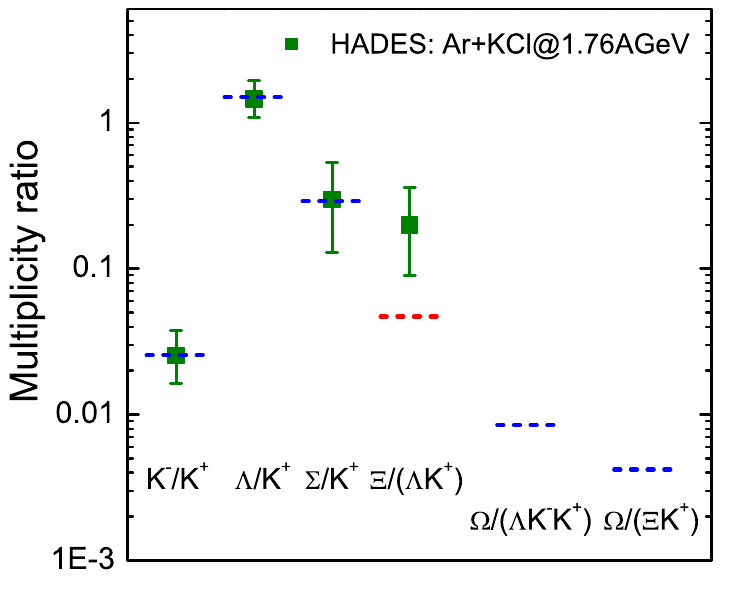}
\caption{
The strange particle ratios (dashed lines) calculated with Eq.~(\protect\ref{S-rat})
in comparison with the available experimental values (symbols).
}
\label{fig:HADES}
\end{figure}

Having the normalization factors and the chemical potential at our disposal, we can calculate the multiplicity ratios of strange particles as functions of the freeze-out density and temperature,
\begin{align}
& R_{K^-/K^+}=\eta\frac{\langle M_{\bar K}^{(1)}+ M_{\bar K}^{(2)}\rangle }{(1+\eta)\, \mathcal{M}_{K^+}} \,,
\qquad \qquad\qquad
R_{\Lambda/K^+} = \frac{1}{\mathcal{M}_{K^+}} \Big\langle M_\Lambda^{(1)} +M_\Lambda^{(2)} +
\eta\frac{M_\Sigma^{(1)} +M_\Sigma^{(2)}}{\eta^2+\eta+1}\Big\rangle \,,
\nonumber\end{align}
\begin{align}
& R_{\Sigma/K^+} = \frac{\eta^2+1}{2(\eta^2+\eta+1)}
\frac{\langle M_\Sigma^{(1)}+M_\Sigma^{(2)}\rangle }{\mathcal{M}_{K^+}} \,,
\quad
R_{\Xi/\Lambda/K^+} = \frac{\frac{\eta}{1+\eta}\langle \big(M_\Xi^{(2)}+M_\Xi^{(3)}\big)\rangle}
{\langle M_\Lambda^{(1)} +M_\Lambda^{(2)} + \eta \frac{M_\Sigma^{(1)} +M_\Sigma^{(2)}}{\eta^2+\eta+1}\rangle
\mathcal{M}_{K^+}} \,,
\nonumber
\end{align}
\begin{align}
& R_{\Omega/\Lambda/ K^{-}/K^+} = \frac{(1+1/\eta)\langle M_\Omega^{(3)}+ M_\Omega^{(4)} \rangle} {\langle
M_\Lambda^{(1)}+ M_\Lambda^{(2)} \rangle \langle M_{\bar{K}}^{(1)}+ M_{\bar{K}}^{(2)} \rangle
\mathcal{M}_{K^+}} \,,
\quad
R_{\Omega/\Xi/K^+} = \frac{(1+1/\eta)\langle M_\Omega^{(3)}+ M_\Omega^{(4)} \rangle } {\langle M_\Xi^{(2)} +
M_\Xi^{(3)} \rangle\, \mathcal{M}_{K^+}} \,.
\label{S-rat}
\end{align}
These relations are applied in Ref.~\cite{KTV12} to the description of multiplicity ratios of strange particles detected by the HADES collaboration in Ar+KCl collisions at $1.76\,A$GeV~\cite{HADES-Xi}. The main effect from the exact strangeness conservation prove to comprise a reduction of the ratio $R_{\Xi/\Lambda/K^+}$ by factor $\sim 0.5$, the ratio $R_{\Omega/\Lambda/ K^{-}/K^+}$ by $\sim 0.2$ and the ratio $R_{\Omega/\Xi/K^+}$ by factor $\sim 0.4$ compared to the results obtained without exact strangeness conservation in each event. The inclusion of in-medium potentials for hyperons and $\bar{K}$ mesons, as discussed in Sect.~\ref{sec:medium} is important, leading to much better agreement with the experiment.
For the $\bar{K}$ potential $U_{\bar{K}}=-75$~MeV the freeze-out baryon density $\rho_{B,{\rm fo}}=0.6\,\rho_0$ and temperature $T_{\rm fo}=69$~MeV are found, which agree well with the parameters obtained from the analysis of the pion production in the same energy range~\cite{Voskre-HIC}. The resulting particle ratios are depicted in Fig.~\ref{fig:HADES} in comparison with the experimental ratios. We see that the calculations are in good agreement with the experiment for the rations of singly-strange particles. The multiplicity of $\Xi$ baryons is, however, strongly underestimated.

\begin{figure}
\centering
\includegraphics[height=6cm]{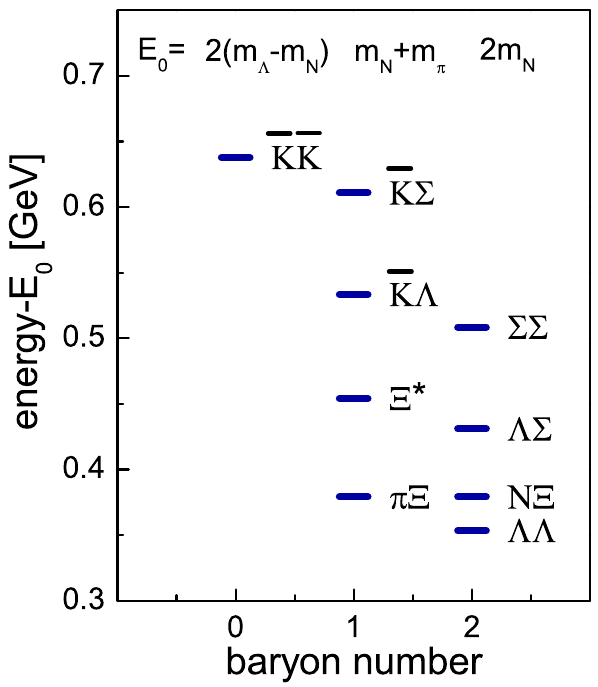}
\includegraphics[height=6.1cm]{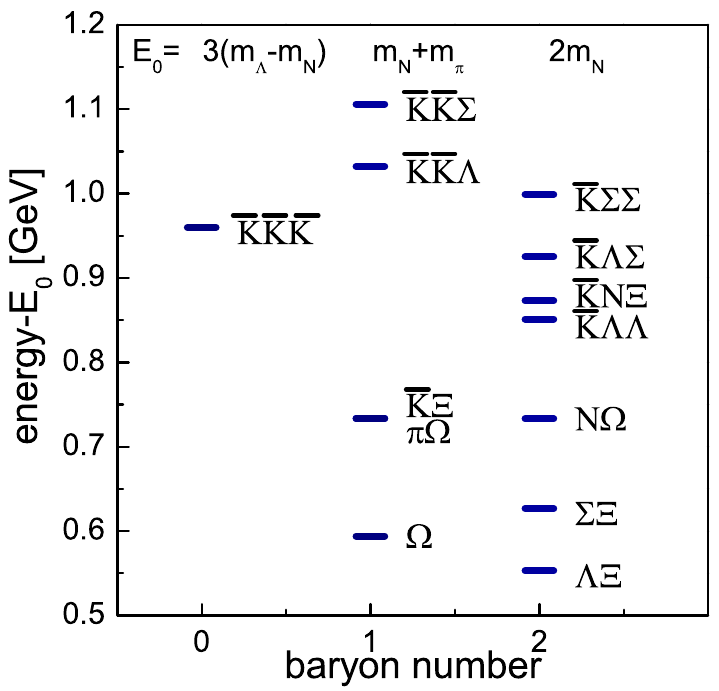}
\caption{
Mass spectrum (energy $E$) of strangeness --2 states (left panel) and --3 (right panel) with baryon numbers 0, 1, and 2 counted from the non-strange ground states with
the energy $E_0$ (shown on top of the figure).}
\label{S-spec}
\end{figure}

Several possible sources of the $\Xi$ enhancement were discussed in Ref.~\cite{KTV12} with the conclusion that to get any substantial increase in the number of $\Xi$'s we have to assume that these baryons are not absorbed after being produced and their number is determined by the rate of direct production reactions, as, e.g., for dileptons.
There two types of reactions: the strangeness creation reactions, like $\bar{K}N\to K\Xi$, $\pi H\to K\Xi$, and strangeness recombination reactions, like $\bar{K}H\to \Xi\pi$ and $HH'\to \Xi N$. The former ones have very high thresholds and, therefore, are not operative at SIS energies.
Strangeness recombination reactions are secondary processes involving two strange particles. They can be promoted statistically. We argue that once several $s$ quarks appear in  hadronic system its more energetically favorable to ``store'' them together in a multistrange state, $\Xi$ or $\Omega$. In Fig.~\ref{S-spec} we show the mass spectrum of strangeness --2 and --3 states in channels with different baryon numbers $B$. We see that in both single and double baryon channels the states with the $\Xi$ and $\Omega$ baryon are at the bottom of the spectrum. This means that $\Xi$s and $\Omega$s would play a role of a strangeness reservoir, being filled with a decrease of the temperature. The only sink is the reaction $\Xi N\to \Lambda\Lambda$ which has, however, a relatively small cross section~\cite{Polinder07}.

\section{Multistrageness at LHC}

We turn now the the highest energies reached in HICs so far. The ALICE collaboration at CERN LHC
reported~\cite{ALICE} the momenta distributions of particles produced in Pb+Pb collisions at $\sqrt{s_{NN}}=2.76\,A$TeV. We analyze these data with the help of the DRAGON Monte Carlo event generator~\cite{dragon}. The DRAGON based on the blast wave model~\cite{blast} but includes not only the particles stable with respect to strong interactions, but also hadron resonancnes, handling altogether 277 hadron species. The particle distribution in the transverse momentum ($p_T$) and rapidity $y$ space is given by the integral $\rmd^2N/\rmd p_T \rmd y=\int\rmd^4 x S(x,p)$ over the source function
\begin{align}
\label{e:Sfun}
S(x,p)\, d^4x = d_i \, \frac{m_t\, \cosh(\eta-y)}{(2\pi)^3}  n_\pm(p_\mu u^\mu - \mu_i)/T)
\theta( 1 -r/R)r\, \rmd r\, \rmd\varphi\, \delta(\tau - \tau_0)\,\tau\, \rmd\tau\, \rmd\eta\,  .
\end{align}
Assuming that the fireball expands dominantly in a longitudinal direction we use here
the longitudinal proper time $\tau = \sqrt{t^2 - z^2}$ and space-time rapidity
$\eta = \frac{1}{2}\ln((t+z)/(t-z))$. A position in the transverse plain is defined by polar coordinates $r$, $\varphi$. The statistical distribution function $n_\pm(x)=1/(\exp(x)\pm 1)$ is used with sign $+$ for fermions and $-$ for bosons, whereas $d_i$ stands for the spin degeneracy of the hadron species. All members of isospin multiplets are treated separately.
The source function (\ref{e:Sfun}) assumes sharp freeze-out along the hypersurface
$\tau = \tau_0$ and uniform density distribution within the radius $R$.
This implies that the freeze-out time does not depend on the radial coordinate.
The fireball expansion is parameterized by the velocity field
$
u^\mu = \big(  \cosh\eta_t \cosh\eta,\, \sinh\eta_t\cos\varphi,  \sinh\eta_t \sin\varphi,\, \cosh\eta_t\sinh\eta\big)
$
where the transverse velocity is such that $ v_t = \tanh\eta_t = \eta_f({r}/{R})^n$\,.
In this relation $\eta_f$ defines the transverse flow gradient and $n$ selects various
transverse velocity profiles. The mean transverse velocity is then equal to
$\langle v_t \rangle = 2\eta_f/(n+2)$\,.
The transverse size $R$ and the freeze-out proper time $\tau_0$ influence total
normalizations of transverse momentum spectra. However, in this study we ignore
those and hence we have no sensitivity to these geometric parameters.

With the help of DRAGON we fitted spectra of pions, charged kaons, protons and anti-protons, together with spectra of $K^0$ mesons and $\Lambda$s for the most central collisions~\cite{ALICE}. The relative particle abundances in the collision are fixed at the moment of the chemical freeze-out characterized by the temperature $T_{\rm ch}=152$~MeV and the baryon chemical potential $\mu_B\simeq 0$~\cite{Milano}. The successive decay of hadronic resonances is explicitly treated in the code. The set of parameters $(T,\eta_f,n)$ in Eq.~(\ref{e:Sfun}) controlling the stage of the kinematic freeze-out of the fireball is optimised for the best agreement with the data. The details of the fitting procedure can be found in~\cite{MT-15}.

\begin{figure}
\centerline{
\includegraphics[width=6cm]{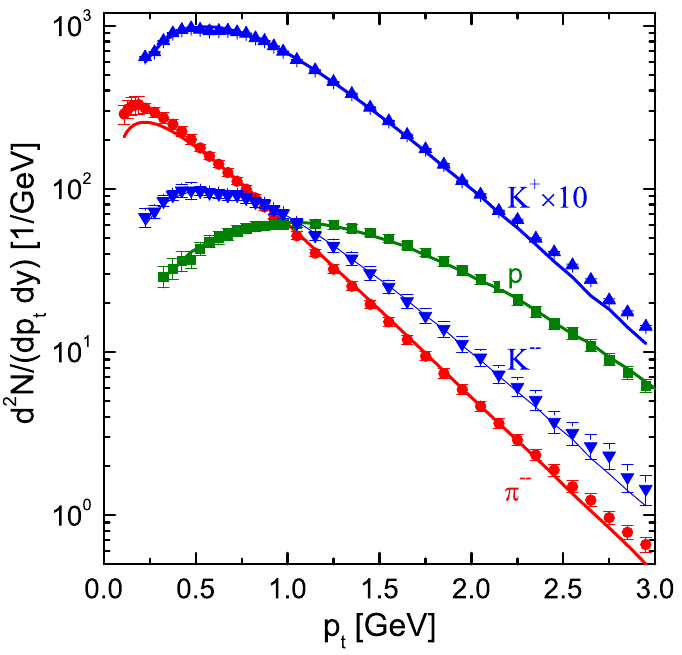}
\includegraphics[width=6cm]{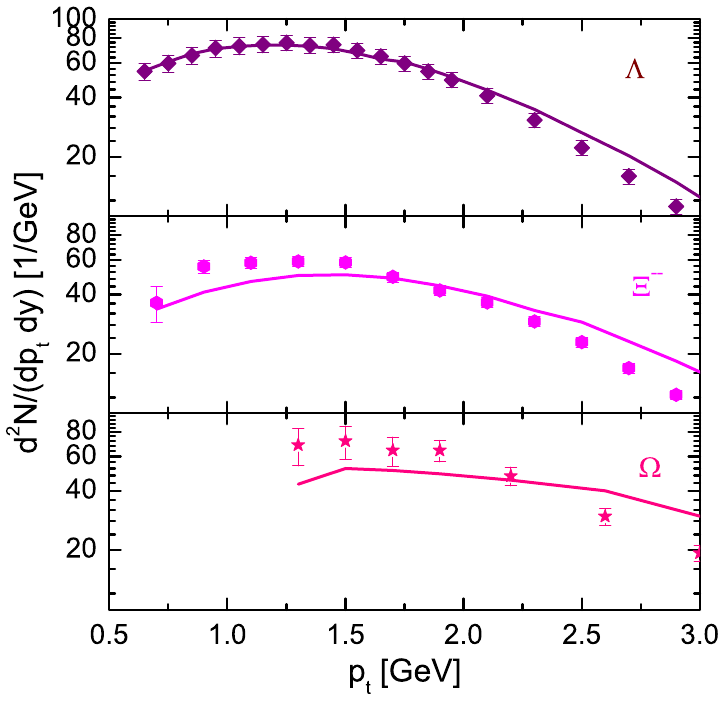}
}
\caption{Transverse momentum distribution of $\pi^-$, $K^\pm$, $p$ (left panel) and
$\Lambda$, $\Xi^-$ and $\Omega$ (right panel) in Pb+Pb collisions at $\sqrt{s_{NN}}=2.76\,A$TeV in comparison with the experimental data~\cite{ALICE}}
\label{fig:LHC}
\end{figure}

For the most central collisions it was found $T=98$~MeV, $\eta_f=0.88$, and  $n=0.69$, that corresponds to the average transverse velocity $\langle v_t\rangle=0.65$. The quality of fits is illustrated in Fig.~\ref{fig:LHC}.  On the left panel we show the spectra of most abundant particles $p$, $\pi^-$, $K^\pm$, which are, in general, very well described by the model in the wide range of transverse momenta.
Only the pion data demonstrate an enhancement at low $p_t$, which cannot be fully accounted for by contributions from decaying resonances. This deficit, we believe, can be resolved by the introduction of non-equilibrium potential for pions.
This would naturally occur if the hadron gas chemically freezes out
at a temperature around 150--160~MeV and then cools down while keeping the
effective ratios of individual species constant. We estimated that the pion chemical
potential at kinetic freeze-out temperature might reach values around 100~MeV.
This is not enough for Bose-Einstein condensation but modifies the spectrum considerably.

The spectrum of $\Lambda$s is also well reproduced as seen in Fig.~(\ref{fig:LHC}), right panel.
However for multi strange particles, which were not included in the common fit, the fit results predict spectra which are in worse agreement with the data.
The experimental data are in general steeper than the predictions. The data would thus indicate lower transverse expansion velocity and earlier freeze-out of multistrange baryons.

\section{Conclusions}

In the present talk we discussed several topics related to the production of particles with several strange quarks.

(i)~For AGS-SPS energies we show that using a hadronic kinetic model, which describes the $K^+$ meson production, we are able to reproduce excitation functions of the multiplicity ratios $K^\pm/\pi^\pm$ and $\Lambda/\pi$. In the same time, the $\Xi/\Lambda$ ratio remains systematically underestimated for AGS-SPS energies.

(ii)~The enhanced yields of $\phi$ mesons at SIS and AGS energies, as we argue, could ask for account of a new type of reactions in which the $\phi$ production is assisted (catalyzed) by a hyperon or a kaon. Such a correlation among $\phi$s and strange particles could manifest itself in the centrality dependence of the $\phi$ yield and in the $\phi$ rapidity distribution.

(iii)~In collisions with a small number of produced strange particle (SIS energies) the exact strangeness conservation in each collision event must be explicitly preserved. This implies for instance that $\Xi$ baryons can be released only in events where two or more kaons are produced.
This has an influence on the observed mean number of $\Xi$s.
The observed enhancement of $\Xi$ production may mean continuous production of
multistrange hyperons when they freeze out right after they are produced in the system.

(iv)~For collisions at the LHC energies we reconstruct a final state of  with the help of DRAGON Monte Carlo generater, based on the blast wave model supplemented with the decay of hadronic resonances. It is found that pion, kaon, (anti)proton and lambda spectra can be satisfactorily described by the common freeze-out temperature and flow parameters. However multistrage baryons seem to decouple at earlier stage of the collision at higher temperatures and weaker transverse flow.

\acknowledgments
The work was supported by the Ministry of Education and Science of the Russian Federation (Basic part).
The work of E.E.K, I.M., and B.T. was partially supported by grants APVV-0050-11, VEGA 1/0457/12 (Slovakia). B.T. also acknowledges M\v{S}MT grant LG13031 (Czech Republic).
Computing was partially performed in the High Performance Computing Center of the Matej Bel University in Bansk\'a Bystrica using the HPC infrastructure acquired in project ITMS 26230120002 and 26210120002 (Slovak infrastructure for high-performance computing) supported by the Research \& Development Operational Programme funded by the ERDF.

\end{document}